**Title:** Towards control of cellular decision-making networks in the epithelial-to-mesenchymal transition

**Authors:** Jorge Gómez Tejeda Zañudo[1,2,3,*], M. Tyler Guinn[4,5,6], Kevin Farquhar[5], Mariola Szenk[4,5], Steven N. Steinway[7], Gábor Balázsi[4,5,*], Réka Albert[1,8,*]

[1]Department of Physics, Pennsylvania State University, University Park, PA 16802, USA
[2]Department of Medical Oncology, Dana-Farber Cancer Center, Boston, MA 02215, USA
[3]Cancer Program, Eli and Edythe L. Broad Institute of Harvard and Massachusetts Institute of Technology, Cambridge, MA 02142, USA
[4]Biomedical Engineering Department, Stony Brook University, Stony Brook, NY 11794 USA
[5]Laufer Center for Physical and Quantitative Biology, Stony Brook University, Stony Brook, NY 11794, USA
[6]Stony Brook Medical Scientist Training Program, 101 Nicolls Road, Stony Brook, NY 11794, USA
[7]Department of Medicine, Johns Hopkins University School of Medicine, Baltimore, MD 21287, USA
[8]Department of Biology, Pennsylvania State University, University Park, PA 16802, USA
*Correspondence: jgtz@phys.psu.edu (J.G.T.Z.), gabor.balazsi@stonybrook.edu (G.B.), rza1@psu.edu (R.A.).

**Abstract:**
We present the epithelial-to-mesenchymal transition (EMT) from two perspectives: experimental/technological and theoretical. We review the state of the current understanding of the regulatory networks that underlie EMT in three physiological contexts: embryonic development, wound healing, and metastasis. We describe the existing experimental systems and manipulations used to better understand the molecular participants and factors that influence EMT and metastasis. We review the mathematical models of the regulatory networks involved in EMT, with a particular emphasis on the network motifs (such as coupled feedback loops) that can generate intermediate hybrid states between the epithelial and mesenchymal states. Ultimately, the understanding gained about these networks should be translated into methods to control phenotypic outcomes, especially in the context of cancer therapeutic strategies. We present emerging theories of how to drive the dynamics of a network toward a desired dynamical attractor (e.g. an epithelial cell state) and emerging synthetic biology technologies to monitor and control the state of cells.

**Main text:**

The epithelial-to-mesenchymal transition (EMT) is essential to the normal embryonic development of the metazoan body, differentiation into tissues and organs, and wound healing (Thiery et al. 2009; Kalluri and Weinberg 2009). In disease, EMT contributes to organ fibrosis (Rastaldi et al. 2002) and the formation of metastases in cancer (Grünert, Jechlinger, and Beug 2003). EMT in various healthy and diseased cells as well as tissues is controlled by a shared core regulatory network interacting with accessory context-specific signaling pathways (De Craene and Berx 2013). The cellular changes induced by EMT, such as enhanced motility and loss of cell-cell adhesion, are also shared between the physiologically normative and pathogenic transitions, especially in metastasis (J. Yang and Weinberg 2008). Below, we briefly summarize the molecular programs and mediators of EMT shared in development, wound healing, and cancer metastasis.

**Regulatory networks of the epithelial-mesenchymal transition**

EMT is usually classified into three major types. Type 1 EMT contributes to embryonic development, particularly blastocyst implantation during early pregnancy and gastrulation (Kalluri and Weinberg 2009; Ohta et al. 2007; Thiery et al. 2009). The Wnt signaling pathway coordinates with the transforming growth factor beta (TGFβ) and fibroblast growth factor (FGF) signaling pathways to induce EMT during gastrulation (Lamouille, Xu, and Derynck 2014; Heisenberg and Solnica-Krezel 2008; Skromne and Stern 2001). Prior to neural crest formation, Wnt and FGF signals induce mesoderm development during gastrulation (Goto et al. 2017). After gastrulation, bone morphogenetic proteins (BMPs) also interact with Wnt and FGF signaling in EMT to form cells in the neural crest (Sauka-Spengler and Bronner-Fraser 2008).

Type 2 EMT is involved in wound healing and organ fibrosis (Burns and Thomas 2010). An important distinction of type 2 EMT is the ability of epithelial or endothelial cells to transition into fibroblast-like cell types which secrete extracellular matrix (ECM) proteins (E. M. Zeisberg et al. 2007; M. Zeisberg et al. 2003). The ECM is composed of polysaccharides, supporting proteoglycans, and fibrous proteins such as collagens, laminins, and fibronectins. The structure of the ECM mechanically supports cells to form tissue and confers mechanical forces and stress signals (Frantz, Stewart, and Weaver 2010). Integrins, adhesive proteins that interact with the ECM, as well as TGFβ signaling mediate type 2 EMT during fibrosis of the lung (Kim et al. 2006). Conversely, BMP7 inhibits EMT in liver and heart fibrosis (E. M. Zeisberg et al. 2007; M. Zeisberg et al. 2003). Cellular markers that identify these fibroblast-like cells in organ fibrosis include fibroblast-specific protein 1 (FSP1), vimentin, and collagen I (M. Zeisberg et al. 2007; Burns and Thomas 2010; Okada et al. 1997). FSP1 is associated with the early development of breast cancer metastasis, highlighting the shared regulatory mediators of EMT in both fibrosis and cancer (Xue et al. 2003).

Type 3 EMT in cancer is typically associated with metastasis (Thiery 2002; Xue et al. 2003; Heerboth et al. 2015; J. Yang and Weinberg 2008). Contrary to type 2 EMT in wound healing, type 3 EMT in cancer was proposed to involve enhanced motility, invasiveness and degradation of ECM during the migration of cells across the basement membrane and their presence at the invasive front of primary tumors (Kalluri and Weinberg 2009). Filopodia, membrane protrusions composed of tightly bundled actin, are essential to this migratory behavior. Cancer cells at the invasive front migrating out of the primary tumor utilize these finger-like protrusions to sense the environment as well as tether the cell to the substrate through cell-ECM adhesion receptors such as integrins (Arjonen, Kaukonen, and Ivaska 2011; Jacquemet, Hamidi, and Ivaska 2015). Myosin-X is a critical regulator of filopodia formation, whose overexpression results in increased cell invasion and whose silencing decreases cell invasion. Mutant p53 drives increased myosin-X expression in cancer cells (Y.-R. Li and Yang 2016). Other components of the cytoskeleton, such as actin stress fibers, are also reorganized during EMT in cancer to facilitate migration and invasion (Yilmaz and Christofori 2009).

Cancer cells that have transitioned into a mesenchymal phenotype may undergo mesenchymal-epithelial transition (MET) to form secondary tumors at distant organs, which then resemble the primary tumor

(Yao, Dai, and Peng 2011; Gunasinghe et al. 2012). Additionally, a partial or hybrid EMT phenotype has characteristics of both the epithelial and mesenchymal states, where cells with the hybrid phenotype can migrate collectively through the extracellular matrix while maintaining cell-cell junctions (Revenu and Gilmour 2009; Friedl and Gilmour 2009; Micalizzi, Farabaugh, and Ford 2010; Zhang et al. 2014; Jolly et al. 2016). During such migration, leader cells at the forefront of migration sense molecular cues through chemotaxis, degrade the matrix (creating tracks for other cells to follow), and drag coupled cells (Aman and Piotrowski 2008; Ilina and Friedl 2009; Rørth 2007). Such highly invasive leader cells can drive non-invasive or poorly invasive cells to invade (Chapman et al. 2014). Besides acting as a medium that invasive cells can degrade, the ECM can transduce mechanical forces, inducing EMT and subsequent metastasis in breast cancer under increasing matrix stiffness (Wei et al. 2015). Therefore, the physical microenvironment also serves as an important mediator of EMT.

Currently, disputes continue about the number and properties of partial EMT states (Jordan, Johnson, and Abell 2011; Schliekelman et al. 2015; Grigore et al. 2016), which may be tumor type- and environment-dependent. In primary mammary and skin tumors, there is evidence of multiple hybrid epithelial/mesenchymal states that differ in expression of distinctive markers and metastatic potency (Pastushenko et al. 2018). Blood-circulating breast cancer cells can have multiple intermediate states that differentially contribute to chemotherapy resistance (Yu et al. 2013). In high-grade serous adenocarcinoma cells, three partial EMT states are classified along an epithelium-mesenchymal phenotypic axis, with SNAIL2 and TCF21 pathway cross-talk and subcellular localization promoting transitions to a state (Varankar et al. 2018). Another controversy refers to the role of EMT in metastasis (Jolly, Ware, et al. 2017). Some recent studies indicate that cancer cells might metastasize without ever undergoing full EMT in pancreatic ductal adenocarcinoma (PDAC) (Zheng et al. 2015) and breast cancer (Fischer et al. 2015). Commentaries and responses to the latter breast cancer study have noted a lack of coverage for multiple EMT markers in the study's genetic tracing system (Diepenbruck and Christofori. 2016; Ye et al. 2017). Moreover, a response to the PDAC study argued that the lack of genetic tracing and inability to ablate all drivers of EMT weaken claims of EMT-independent PDAC metastasis (Aiello et al. 2017). Adding to this controversy, single cells and collective cell clusters in partial EMT states were reported to directly contribute to metastases in vivo (Puram et al. 2017; Aiello et al. 2018). In summary, the contribution of EMT in general and its intermediate states in particular in promoting metastasis is under increasing scrutiny; a thorough investigation of additional EMT drivers and markers in different cancers may clarify this relationship.

Although all three types of EMT rely on context-specific signals for induction, a shared core regulatory network seems to govern the ultimate cellular decision to undergo EMT. This network incorporates sets of transcription factors and microRNAs that mutually repress each other (Figure 1). While many signaling pathways can induce EMT, including tyrosine kinase receptor pathways such as MAPK (Doehn et al. 2009), platelet-derived growth factor (PDGF) signaling (L. Q. Yang, Lin, and Liu 2006), hepatocyte growth factor signaling (Grotegut et al. 2006), and epidermal growth factor signaling (Lo et al. 2007), TGFβ signaling is the most well-characterized EMT inducer (Jian Xu, Lamouille, and Derynck 2009; Lamouille, Xu, and Derynck 2014). TGFβ mediates the earliest step in this core EMT regulatory network by increasing SNAIL1, SNAIL2, and TWIST expression (Thuault et al. 2006).
The transcriptional repressors SNAIL1 and TWIST then inhibit E-cadherin expression, leading to loss of cell-cell junctions and enhanced motility (Batlle et al. 2000; Cano et al. 2000; Lamouille, Xu, and Derynck 2014). In an opposing role, the microRNA inhibitor of SNAIL1, miR-34, is in turn inhibited by SNAIL1 and ZEB1, creating a double negative feedback loop between the SNAIL1 transcriptional repressor and the inhibitory microRNA (Siemens et al. 2011; Ahn et al. 2012). SNAIL1 can also repress its own promoter, leading to negative auto-regulation of SNAIL1 expression (Peiro et al. 2006). Additionally, TWIST cooperates with SNAIL1 and ETS1 to induce the expression of the pro-EMT transcriptional repressor ZEB1 (Dave et al. 2011). As in the first layer, ZEB1 and the miR-200 family of microRNAs repress each other in a double negative feedback loop, with miR-200 acting as an EMT inhibitor (Burk et al. 2008; Bracken et al. 2008). In fact, miR-200 is down-regulated in several malignancies including breast and bladder cancer (Wiklund et al. 2011; Hayes, Peruzzi, and Lawler 2014). SNAIL1 also inhibits miR-200 expression, though with lower inhibitory strength than ZEB1 (Burk et al. 2008). In contrast to SNAIL1's negative auto-regulation, ZEB1 indirectly activates its own expression through self-sustaining regulation of splicing for CD44, which then activates ZEB1 (Preca et al. 2015).

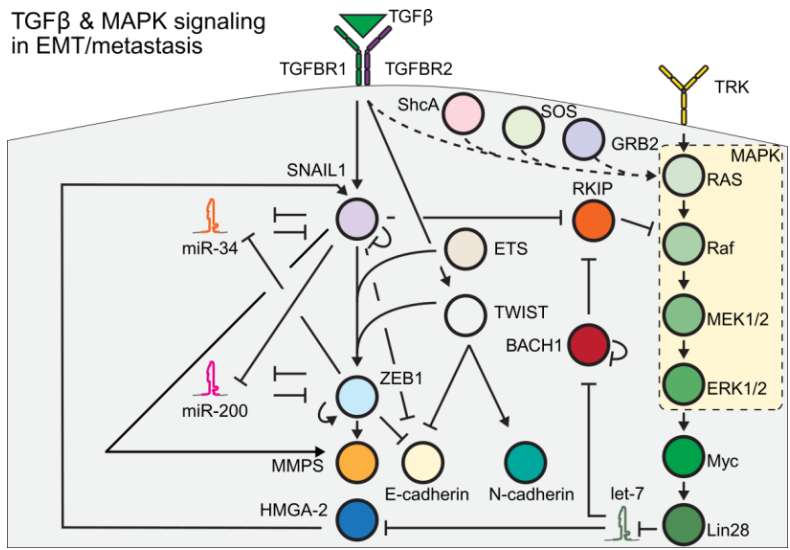

**Figure 1: The TGFβ-induced EMT/metastasis network includes a core regulatory network as well as multiple cross-talks and feedback loops.**
TGFβ mediated EMT drives the transcription factors SNAIL1 and TWIST to induce the first layer of the core regulatory network. ZEB1 is then induced by SNAIL1 and TWIST, which jointly leads to loss of E-cadherin expression, increased expression of N-cadherin (mesenchymal marker), and an increased expression of matrix metalloproteinases (MMPs), which promote invasiveness and metastasis. MicroRNAs act to repress these transcription factors for preventing EMT, which are themselves repressed by SNAIL1 and ZEB1. The MAP Kinase pathway, mediated by tyrosine receptor kinases (TRK), interacts with the core network to indirectly increase SNAIL1 expression. This pathway also participates in a feedback loop with BACH1 and RKIP. Circles represent proteins and transcription factors. MicroRNAs are indicated by stem-loops. Connecting edges between intermediates indicate co-facilitation of an interaction while dashed interactions are composed of multiple sequential interactions. The MAPK pathway is designated by the dashed box in the cytoplasm.

Additionally, the core regulatory network cross-talks with the MAPK pathway through phosphorylation of SHCA (Figure 1), which indirectly drives RAS activation (M. K. Lee et al. 2007). Moreover, the MAPK pathway indirectly inhibits the tumor suppressing microRNA let-7, which inhibits the pro-metastasis BACH1 transcriptional repressor and the SNAIL1-activator HMGA2 (Y. S. Lee and Dutta 2007; Yun et al. 2011). RAF kinase inhibitory protein (RKIP) serves as an anti-metastatic inhibitor of MAPK that ultimately decreases the expression of BACH1 though increasing let-7 expression (Dangi-Garimella et al. 2009). In turn, RKIP is directly repressed by BACH1, forming another double negative feedback loop (Figure 1) adjacent to the core network (J. Lee et al. 2014). Overall, the transcriptional regulators that lead to EMT are shared in development, wound healing, and can be perturbed during cancer progression – but their exact role in metastasis remains to be analyzed. In particular, it is still unclear how exactly these EMT transcriptional regulators are coupled with other toggle switch modules more clearly involved in metastasis, such as RKIP-BACH1 (Yun et al. 2011; J. Lee et al. 2014; Kolch et al. 2015).

### *In vitro* models of EMT and metastasis

Phenotypic cues of EMT and cancer metastasis include: cell-cell dissociation, wherein cells lose cell-cell adhesion and migrate away from a primary tumor; substrate degradation, wherein cells invade the basement membrane; and cellular migration and intravasation across the endothelium into blood and lymphatic vessels, followed by circulation throughout the body, extravasation, and colonization, the formation of secondary tumors in areas other than the primary organ. Due to the complexity of these processes and their dependence upon the tumor microenvironment, the metastatic capabilities of cells have predominantly been examined through *in vivo* mouse models, such as tissue grafting or tail vein injection (S. Yang, Zhang, and Huang 2012). However, such *in vivo* models are costly, slow, and inherently limited in imaging capability. These systems do not allow efficient probing of mechanical features of metastasis or of individual rate limiting steps of the metastatic process. Likewise, *in vivo* models do not lend themselves to rapid experimental manipulations and parameter control or prototyping of therapeutics. Lastly, murine models do not accurately mimic human tumorigenesis (Rangarajan et al. 2004). Thus, *in vitro* reductionist approaches are needed to accelerate discovery and to adequately understand the molecular mechanisms of human tumor metastasis.

*In vitro* EMT and metastasis models vary in their ability to recapitulate *in vivo* conditions. Cellular migration, growth, morphogenesis, differentiation, drug sensitivity, and gene expression exhibit differences depending on whether they are studied in 2- or 3-dimensional systems (Yamada and Cukierman 2007; Weigelt et al. 2010). Here we will give a brief overview of various model systems that have been developed for the study of EMT and metastasis.

Changes in ECM proteins as well as in the expression level and activation state of integrins are common features of cancer progression and metastatic potential (Bendas and Borsig 2012; Ioachim et al. 2002; Liapis, Flath, and Kitazawa 1996; Rolli et al. 2003; Putz et al. 1999; Zutter, Sun, and Santoro 1998; Lipscomb et al. 2005; D.-M. Li and Feng 2011). Therefore, 2-dimensional (2D) adhesion assays are generally used to initially characterize the functional properties of metastatic cells. In such assays, cells are seeded onto culture plates coated with ECM proteins of the relevant tumor type and metastasis stage under study. Non-adherent cells may then be washed off or detached through shear (Boettiger 2007). These short-term assays can monitor changes in adhesion receptors through the addition of function-blocking antibodies or inhibitors and quantified through counting of fluorescent/stained cells.

Whether it is due to gradients of growth factors, chemokines, adhesive substrates, or matrix stiffness, cellular migration is a key component of the metastatic process. 2D *in vitro* models that assay migratory capacity include wound healing assays, exclusion zone assays, and transwell migration assays. Wound healing or scratch assays measure the migration of cultured cells as they close a gap in a confluent monolayer generated by scratching the surface of culture dish. Cells are typically serum-starved for 8-24 hours prior to the assay to limit the effects of cell turnover on the rate of wound healing. These assays are widely used as they produce time-lapse data, are inexpensive, convenient, and do not require specialized reagents. However, they are difficult to standardize and are not easily reproducible. To remedy inconsistencies in user generated wound size, exclusion, or gap closure, assays generate a cell free zone by culturing cells around a blockade that is then removed. In contrast, transmembrane migration assays utilizing a Boyden chamber do not require cells to be adherent. Cells are seeded in serum-free media in the upper chamber and allowed to chemotactically transmigrate across a semipermeable membrane toward serum-rich media in the lower chamber. These assays enable testing the effects of modifications such as changing the membrane pore size, coating the membrane in ECM components, as well as adding or changing the chemoattractants present in the media. They offer a snapshot of the transmigration process as cells crossing the interface are counted after a set amount of time. Such assays are inherently limited in study time as the concentrations in the two chambers equilibrate over time. As an alternative, the collagen dot assay can overcome this limitation and investigate the role of therapeutic drugs in high throughput over a longer time period (Alford et al. 2016). In this approach, cells in a collagen mixture are dotted onto 96-well plates and migrating cells at the collagen-cell interface are counted after a longer incubation period.

3-dimensional (3D) in vitro models of metastasis can mimic aspects of the tumor microenvironment by combining one or more cell types and molecular factors into a controlled system (Provenzano et al. 2010; Griffith and Swartz 2006). The earliest 3D models were produced through tissues harvested *in vivo* and explanted for short term study (Gähwiler et al. 1997). Due to the lack of vascularization within the model platform, such tissues had to be sufficiently thin to allow for adequate oxygenation and nutrient delivery into the interior. Most 3D models, however, are established from isolated cells from either cell lines or tissues, which have been propagated in culture prior to implantation in a 3D matrix. Such 3D scaffolds are produced from either collagen and other glycoproteins, proteoglycans, synthetic biomaterials, engineered tissues, or even ECM components extracted from living tissue, mimicking cross-linked networks of collagen present in the stromal environment (Sabeh, Shimizu-Hirota, and Weiss 2009; Hutmacher et al. 2010). The availability of different 3D matrices is both a pro and a con, as matrix components may be carefully chosen to simulate different conditions (such as scaffold stiffness) and test hypotheses, but such variability may confound the growing corpus of data on cellular aspects of metastasis. It is therefore extremely important to tailor the *in vitro* model to the research question.

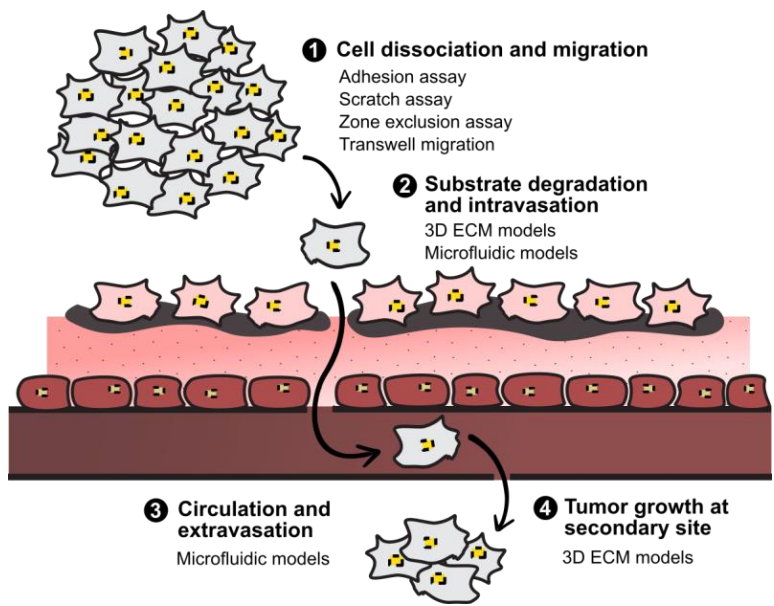

**Figure 2: *In vitro* models investigate various steps of the metastatic cascade.**
Cellular dissociation from the primary tumor and subsequent migration can be studied through adhesion assays, and migratory assays such as wound healing and zone exclusion assays. Degradation of the basement membrane as well as translocation across the ECM is readily recapitulated in 3D cultures. Intravasation and extravasation, wherein cells must squeeze through confined spaces such as gaps in endothelial cell walls, and circulation lends itself to study through microfluidic models. The growth of cancer cells at secondary sites may be recapitulated through 3D cultures such as spheroids.

3D models may recapitulate certain aspects of *in vivo* tumorigenesis that are not possible in 2D culture, such as heterogeneous tumor environment architecture, hypoxia and nutrient availability, spatial organization, cell-cell or cell-matrix interactions, and mechanical forces such as growth induced solid stress (Jain, Martin, and Stylianopoulos 2014). The dynamics of cancer cell aggregates called spheroids are commonly assessed in 3D culture. These spheroids may be composed of many different cell types, including tumor, stromal, and immune cells. Such mixtures are uniquely poised to mimic tumor heterogeneity and morphology by combining cells with differential growth rates, surface-exposed and nutrient-starved cells, hypoxic cells and well-oxygenated cells into a single system (Kunz-Schughart et al. 2004; Frieboes et al. 2006). Likewise, 3D models are particularly apt for monitoring single-cell or collective invasion through ECM substrate degradation by matrix metalloproteinases (Yamada and Cukierman 2007; P. Lu et al. 2011).

Microfluidic models of metastasis have gained traction in recent years due to their ability to bridge the divide between traditional 2D assays and 3D approaches (Bersini et al. 2014). These microfluidic models consist of micrometer-scaled channels and structures often fabricated into polydimethylsiloxane (PDMS) chips in 3 dimensions. Cells are seeded onto such chips and studied under long term time-lapse imaging. As a result, microfluidic models can mimic the tracts and vessels that cancer cells must traverse in order to successfully metastasize. Work with a 3D-confined microfluidic motility assay (Irimia and Toner 2009) suggests that microchannel confinement alone is sufficient to induce collective directional migration in cancer cells. Likewise, through fabricated branched microchannel structures of varying sizes, microfluidic models can be used to determine the velocity and dynamics of cellular migration under varying vascular morphology (Huang et al. 2013). Such devices also lend themselves to the study of cellular co-cultures, or to separate out non-migratory and migratory subsets of heterogeneous tumor populations (Chen et al. 2015).

Microfluidic model systems are not limited to the study of migration and circulation, but may also be used to establish and maintain spatial gradients of chemokines, growth factors, or chemotherapeutic drugs (Liu et al. 2009; Lin et al. 2017). Recent technology mimics the tumor-vascular interface by combining microfluidic chips with 3D ECM gels (Zervantonakis et al. 2012; Liu et al. 2009) and/or endothelial cell boundaries (Zhang, Q. et al. 2012; Chen et al. 2013; Riahi et al. 2014). These 3D microfluidic models consist of independently controllable microchannels connected to 3D ECM hydrogels, allowing precise control of growth-factor gradient and the observation of intravasation or extravasation events through live imaging. Experiments coupling vascular boundaries with endothelial cells have shown that metastatic cancer cells are capable of significant morphological deformation during trans-endothelial migration and thus do not need to disrupt endothelial cell-cell junctions (Chen et al. 2013; Riahi et al. 2014).

**Genetic control by EMT modifiers**

Studies of EMT as a pathologic process in malignancy development has revealed important EMT regulators, including transcription factors and microRNAs in several tissue types of origin (Beerling et al. 2016; Fischer et al. 2015; Krebs et al. 2017; Ye et al. 2017; Zheng et al. 2015). Abnormal intracellular levels of such EMT regulators can drive or suppress EMT, with potential implications for metastasis (Harper et al. 2016; Lambert, Pattabiraman, and Weinberg 2017; Krebs et al. 2017). In the following we define EMT modifiers as tools that change the genetic sequence of one of the components of the EMT regulatory network, such as gene deletion, overexpression, or genome editing. Empirical studies over the last decade have furthered the knowledge foundation of EMT by generating model systems with both knockout (KO) or knock-in (KI) genes for analyzing the necessity of specific biological factors for induction or blocking of EMT or MET (Diaz-Lopez et al. 2015; Brabletz et al. 2017; Krebs et al. 2017; H. J. Lee et al. 2016). With the availability of new genetic engineering tools, such as Cas9, producing over-expression as well as knock-out cell lines and model organisms has become more and more standardized, allowing probing of single-gene deletion or over-expression (Figure 3A). Such studies have provided evidence for the role of EMT in metastasis and the necessity or absence of specific genes and their RNA or protein products in both *in vitro* and *in vivo* models. Isolated perturbation of single EMT-regulator genes allows greater understanding of the underlying genetic network that drives EMT (Weitzenfeld, Meshel, and Ben-Baruch 2016; Spaderna et al. 2008; Werden et al. 2016; Jia et al. 2017).

The experimental methodologies have included over-expression, deletion, or down-regulation of genes such as ZEB1, TWIST, SNAIL1, and FOXC2 in non-small cell lung cancer, breast, renal and colorectal cancer cell lines (H. J. Lee et al. 2016; Cano et al. 2000; Jia et al. 2017; Werden et al. 2016). At a higher-level analysis, there have been *in vivo* studies studying the connection between metastasis and EMT in athymic nude (BALB/c nu/nu) mouse models (Werden et al. 2016; Spaderna et al. 2008). To complement genetic modifications of cell-lines and mouse models for inducing or blocking EMT, other groups have induced EMT with small molecules including TGFβ, TGFα, and EGF in *in vitro* assays, where cells can be easily tracked (Okada et al. 1997). On the other hand, the corresponding mouse models have been more cumbersome to investigate than cell cultures (Fischer et al. 2015; Zheng et al. 2015). Additionally, small molecules such as microRNAs have begun to be utilized for controlling certain cell phenotypes, but delivery of such payloads for *in vivo* models have remained problematic (Dowdy 2017; Cuellar et al. 2015).

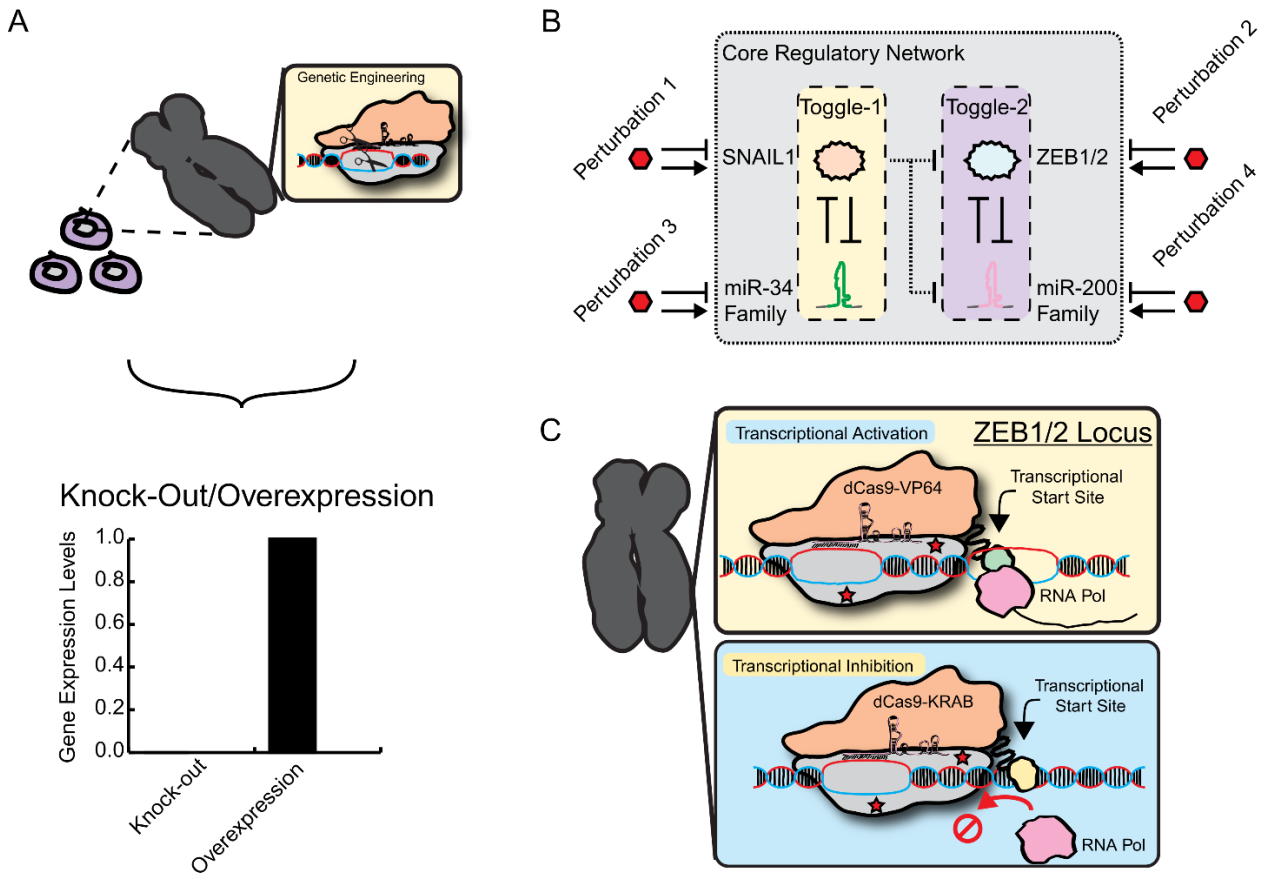

**Figure 3. Synthetic biological EMT modifiers.**
(A) Schematic illustration of a genetic engineering tool example (e.g. Cas9) and outcomes for knocking out or knocking in a specific EMT gene of interest. (B) Biological schematic of the core regulatory network controlling epithelial-mesenchymal transition. Network control via perturbing each toggle switch as critical areas with synthetic microRNAs or anti-sense oligonucleotides. (C) Direct perturbation of an endogenous EMT gene locus with transcriptional tools such as dCas9-VP64 or dCas9-KRAB.

To complement genetic modifying instruments that could be integrated into various host genome locations, biochemical tools could also target known genes of interest (Figure 3B). Various biochemical regulators could be engineered into the genome, such as synthetic microRNAs and anti-sense oligonucleotides. To supplement the permanent changes in genetic modifications of knock-in/knock-out models and external perturbations by chemical means, internal gene modifiers can be introduced into cells to regulate gene expression in a more nuanced manner. For example, transcriptional regulators of EMT genes acting at endogenous sites can be introduced to induce or repress gene transcription using genetic engineering tools such as modified Cas9 proteins (Figure 3C). This can be accomplished by fusing of DNA binding domains (e.g. Cas9+gRNA) to transcriptional activators (e.g. VP64) or transcriptional inhibitors (e.g. KRAB). Using genome engineering tools such as Cas9 and dCas9 adds to the versatility of methods that can target and modify any gene relevant to EMT.

While powerful in elucidating information on endogenous networks regarding EMT, many of the above techniques have the limitations of eliciting 'digital' or 'binary', often irreversible responses (Figure 3A). These methods are useful for analyzing the extremes of gene expression, such as complete shut-down or maximal expression, but are limited in exploring dynamic gene expression levels in endogenous biological settings. Considering the potential relevance of EMT or metastasis-regulator levels (not just their presence/absence), it will be important to design methods to detect and control their expression level dynamically in single cells.

**Mathematical models of the molecular networks underlying EMT and metastasis**

Several reviews have covered different aspects of mathematical modeling of EMT (Jolly and Levine 2017; G.T. Zañudo et al. 2018; Jolly, Boareto, et al. 2017; Jolly et al. 2015; Zhang, Tian, and Xing 2016). Here we review a selection of models with a focus on those that highlight the connection between network motifs and feedback loops with the EMT process and hybrid epithelial/mesenchymal (E/M) states.

*Ordinary differential equation models*

Mechanistic mathematical models based on ordinary differential equations (ODEs) allow a quantitative understanding of how the network of interactions and reactions and their underlying kinetics shape the behavior of the system. Several ODE models have been developed in the context of EMT and metastasis (C. Li and Balazsi 2018; M. Lu et al. 2013; Zhang et al. 2014; Jia et al. 2017; Tian, Zhang, and Xing 2013; Zhang et al. 2018; Hong et al. 2015; J. Lee et al. 2014; Jolly et al. 2016). The work of (Tian, Zhang, and Xing 2013) and (M. Lu et al. 2013) were the first to propose that hybrid E/M states could arise because of the double negative feedback loops between SNAIL1 and its microRNA regulator miR-34, and ZEB1, and its microRNA regulator miR-200, respectively. The mechanism for the appearance of such a hybrid state differs between the two models: in the model of (Tian, Zhang, and Xing 2013) both feedback loops are bistable, and the hybrid state corresponds to high SNAIL1/low miR-34 and low ZEB1/high miR-200 activity, while in (M. Lu et al. 2013) the hybrid state is a consequence of tristability in the ZEB1/miR-200 feedback loop. An important difference between these models is that the model of (M. Lu et al. 2013) includes self-activation of the transcription factor ZEB1; this is the underlying reason for its tristability (Huang et al. 2007; Huang 2013). Another difference between the two models is the inclusion by Tian et al. of the autocrine production of TGFβ, which is inhibited by miR-200, and thus forms a net positive loop with SNAIL1 and ZEB1. Current experimental evidence is consistent with hybrid states having different mechanisms that vary between cells and contexts, and has provided support for the mechanism of both models (Jia et al. 2017; Zhang et al. 2014; Hong et al. 2015).

The effect of various transcription factors and signaling pathways on EMT through influencing the SNAIL1/miR-34 and ZEB1/miR-200 loops has been explored in several ODE models. (Zhang et al. 2014) explored using the model in (Tian, Zhang, and Xing 2013) how an intermediate dose of exogenous TGFβ can result in a reversible E-E/M transition by the activation of only SNAIL1, while a high dose of exogenous TGFβ leads to an irreversible E-M transition by the activation of both SNAIL1 and ZEB1. The predictions of the model were verified experimentally using single-cell flow cytometry in the breast cell line MCF10A. Follow up work in (Zhang et al. 2018) explored through modeling and experiments in multiple breast cancer cell lines how SNAIL1 can decode the duration of a TGFβ stimulus through the use of intermediary feedforward loops involving the SMAD and GLI signaling pathways. (Hong et al. 2015) used the model of (Tian, Zhang, and Xing 2013) and included OVOL2, which forms a double negative feedback loop with ZEB1 and inhibits autocrine TGFβ. The addition of OVOL2 increased the system's attractor repertoire by an additional hybrid E/M state, and illustrates how the inclusion of additional feedback loops in the model can result in multiple hybrid states. (Hong et al. 2015) tested their model by using single-cell flow cytometry and analyzing the response of multiple breast cell lines to TGFβ induction or OVOL2 overexpression, which showed consistency with the four steady states predicted by their model.

(Jolly et al. 2016; Jia et al. 2015) tested the effect of coupling additional components and feedback loops to the SNAIL1/miR-34 and ZEB1/miR-200 core. In particular, they found that mutual inhibition between ZEB1 and OVOL, GRHL2, or miR-145, respectively (Jolly et al. 2016; Jia et al. 2015) increased the range of parameter values under which a hybrid E/M state is available, and thus, could be used to increase the stability of hybrid E/M states. The predictions of their model were tested experimentally in a lung cancer cell line that displayed a hybrid E/M phenotype; the model predicted and the experiments confirmed that the hybrid E/M phenotype would be lost in case of knockout of OVOL2 or GRHL2, resulting in an M phenotype. (Boareto et al. 2016; Bocci et al. 2017) found that Notch-Jagged cell-to-cell signaling allowed the stabilization of the hybrid E/M state and the formation of spatial clusters of hybrid E/M states. Consistent with the predictions in (Boareto et al. 2016; Bocci et al. 2017), Jagged was found to be upregulated in tumor cell clusters (Cheung et al. 2016), and its knockdown caused a reduction in cluster formation (Bocci et al. 2018).

Most ODE models of EMT and metastasis have focused on a transcription factor core composed of ZEB1 and SNAIL1 because of their known importance in inducing EMT. Recent experimental work questions whether EMT and the traditional ZEB1 and SNAIL1 transcription factor core are required for metastatic transitions (Fischer et al. 2015; Zheng et al. 2015). Consequently, the search is on for regulatory networks closer associated with metastasis rather than EMT (Yun et al. 2011; J. Lee et al. 2014). (J. Lee et al. 2014) identified a double-negative feedback loop between the TF BACH1 and the metastasis suppressor RKIP in breast cancer. They showed using mathematical modeling and single-cell experiments that the BACH1/RKIP feedback loop supports bistability and that this feedback loop can act independently of the EMT TF SNAIL1. Building on this work, (C. Li and Balazsi 2018) developed an ODE model that coupled BACH1 and RKIP with EMT and stemness TFs/miRNAs. The coupled model had as steady states an antimetastatic E state (low BACH1, low ZEB1), a metastatic M state (high BACH1, high ZEB1), and two hybrid states: an anti-metastatic E/M state (low BACH1, intermediate ZEB1) and a metastatic E-like state (high BACH1, low ZEB1). Thus, this model predicts that BACH1/RKIP can couple with the EMT TFs to create a metastatic M state, but can also act independently of the EMT TFs and give rise to a hybrid metastatic E-like state. This model also predicts that the multi-stability due to the multiple double-negative feedback loops gives rise to intermediate cellular states accessible from the antimetastatic E state. This lowers the robustness of the antimetastatic E state and amplifies the nongenetic heterogeneity of gene expression, suggesting a role for gene expression noise in EMT fate determination and metastasis (Li and Balazsi, 2018).

*Discrete dynamic models*

In contrast to the continuous state variables in ODE models, discrete dynamic (also called logical) models assume a discrete number of states for each node. Discrete models lack quantitative aspects that ODE models excel at (e.g. dose-response curves or parameter bifurcation diagrams), and instead focus on a coarse-grained view of the dynamics of the system and how the logic of the interactions give rise to a model's repertoire of dynamical behaviors (attractors). One of the methods to determine a logical model's repertoire is to identify self-sustaining positive feedback loops called stable motifs. These stable motifs have an associated state which determines a trap subspace in the system's state space: after the stable motif attains the associated state, the system cannot escape the trap subspace (Jorge G. T. Zañudo and Albert 2013; Gan and Albert 2018; Zanudo and Albert 2015). Several discrete network models of EMT have been developed (Steinway et al. 2014, 2015; Cohen et al. 2015; Khan et al. 2017; Udyavar et al. 2017; Wooten and Quaranta 2017; Méndez-López et al. 2017). Some of the text in this subsection is based on a recent review of discrete dynamic models in cancer by a subset of the authors (G.T. Zañudo et al. 2018).

One of the initial efforts was by Steinway et al., who constructed a logical model of EMT in the context of hepatocellular carcinoma (Steinway et al. 2014, 2015). The network model includes E-cadherin (CDH1) as an EMT marker, seven known EMT transcriptional regulators (including SNAIL1, SNAIL2, ZEB1, ZEB2, HEY1 and TWIST), a microRNA (miR-200), multiple extracellular signals (e.g. Wnt, Shh, TGFβ), and diverse signaling pathways (PI3K, MAPK, and receptor tyrosine kinases). The model predicted that EMT is a robust outcome, induced by a variety of external and internal signals, many of which sustain each other (e.g. TGFβ concurrently induces the Wnt and Shh signaling pathways), a prediction which was validated experimentally in human and mouse hepatocellular carcinoma cell lines (Steinway et al. 2014). In follow up work (Steinway et al. 2015) used the model and a stable-motif-based network control approach (Zanudo and Albert 2015) to predict, and cell line experiments to confirm, that the combinatorial inhibition of SMAD and proteins in the RAS or NOTCH pathways but not SMAD alone, can suppress TGFβ-driven EMT. (Steinway et al. 2015) also predicted that certain perturbations (e.g. SMAD knockout) give rise to hybrid states intermediate between the epithelial and mesenchymal states, thus emphasizing the need for combinatorial therapies to fully suppress EMT and invasion. The hybrid states identified in (Steinway et al. 2015) are characterized by a high activity of several transcription factors associated with EMT (SNAIL1/2, ZEB1/2, and TWIST) and low activity of the EMT transcription factor HEY1 and of signaling pathways typically associated with EMT (RAS, PI3K/AKT, NOTCH, SHH, and WNT). Initial model construction only yielded epithelial or mesenchymal states. Only once computational perturbations were employed (knockouts and overactivation), did the hybrid states become apparent.

(Cohen et al. 2015) constructed a logical model of signaling and regulatory mechanisms leading to EMT, invasion, migration and metastasis from a curated selection of pathways (e.g. Wnt, Notch and PI3K-AKT), known EMT markers and drivers (CDH1/2, VIM, SNAIL1/2, ZEB1/2, and TWIST), microRNAs (miR-34 and miR-200), and p53 family members. The network model was iterated to be able to recapitulate the effect on EMT/invasion/migration/metastasis of most mutants in a set of more than 15 *in vitro* and *in vivo* studies. Among the model predictions are that the Notch gain-of-function and p53 loss-of-function double mutant is the most efficient in inducing metastasis and that the TGFβ pathway is required for metastasis in this mutant.

(Udyavar et al. 2017) constructed a network model of transcription factors (TFs) for the neuroendocrine (epithelial-like) and non-neuroendocrine (mesenchymal-like) phenotypes in small-cell lung cancer (SCLC) using a mixed bioinformatics and literature-based approach. The model predicted the existence of neuroendocrine-like and non-neuroendocrine-like attractors, a prediction that was confirmed experimentally in a subset of the cell lines and tumors using a selection of the TFs characterizing the phenotypes. The model did not exhibit hybrid attractors and thus was unable to predict the hybrid phenotypes observed experimentally in some cell lines and tumors.

(Méndez-López et al. 2017) developed a Boolean network model of the gene regulatory and cell signaling processes underlying senescence, inflammation, and EMT in carcinomas (cancers of epithelial origin). The model reproduced the expected gene expression state of epithelial, senescent, and mesenchymal phenotypes as attractors, reproduced (through perturbation-driven transitions between phenotypes) the commonly observed progression by which epithelial cells first become senescent and then acquire a mesenchymal stem–like phenotype, and predicted that the transition into the mesenchymal phenotype is accelerated by inflammation, consistent with the poor prognosis of pro–inflammatory conditions.

*Continuous and discrete models complement each other*

As true in general, the continuous and discrete models of EMT are complementary of each other (Calzone, Barillot, and Zinovyev 2018). The continuous models usually describe each node at a more detailed resolution, for example they describe how a microRNA (e.g. miR-200) can bind in multiple sites to the mRNA of its target TF (e.g. ZEB1) and form various mRNA-microRNA complexes. Yet, as apparent from our summary of the key results of ODE models of EMT, these key results can often be summarized by invoking a small number of qualitative states at the level of network motifs (e.g. a bistable switch based on mutual inhibition between ZEB1 and miR-200). A discrete model that uses these qualitative states is thus representative of the key features of the continuous model.

It is likely to be generally true that discrete models correspond to coarse-grained versions of continuous models that are mechanistic at the elementary reaction level (Calzone, Barillot, and Zinovyev 2018). The nodes of the discrete model are coarse-grained representations of modules or motifs of the continuous system, and the discrete states represent the attractors of these modules/motifs. Thus, the number of relevant states of a node (e.g., as defined by the number of steady states) in the continuous model would inform the states and regulatory functions of the discrete model; due to the gain in simplicity characterizing each node, a discrete model can have a more system-level scope. Such a model can be customized to fit cell-type specific networks, for example, nodes not actively expressed in a certain cell can be removed from the model. It can evaluate which features of the network (elements, pathways, feedback loops) are most informative. Based on the insights gained, it can suggest the most promising avenues to expanding the continuous model. Ultimately, a most useful use case of a well-parametrized model may be a hybrid of continuous and discrete models with both detailed and coarse-grained nodes, which allows us to zoom-in or zoom-out depending on the question being asked.

**Network models predict that multiple transcription factors and signaling pathways can underlie hybrid E/M states**

Most continuous models assume that the decision of whether cells are epithelial, mesenchymal, or in an intermediate state is made by a transcription factor (TF) regulatory core that involves the main families of TFs known to be sufficient for (in other words, known to drive) EMT, namely, SNAIL and ZEB. These core TFs interact with each other in the decision-making process of EMT, downregulate the transcription of epithelial genes (e.g. E-cadherin and claudins), and upregulate mesenchymal genes (e.g. Vimentin and N-cadherin). In addition to the SNAIL and ZEB families, additional TFs known to drive EMT are often included in discrete models, with some examples being TWIST, FOXC2, and HEY1 (Lamouille, Xu, and Derynck 2014; Héctor Peinado, Olmeda, and Cano 2007; R. Weinberg 2013; De Craene and Berx 2013; Jian Xu, Lamouille, and Derynck 2009). An implicit assumption of using these TFs (or a subset) as the underlying EMT TF regulatory core, and which should be used with caution, is that other TFs that drive EMT either do it through this TF core or are parallel to the TF core but act similarly to it.

In reality, we know that several signaling pathways and transcriptional programs can induce EMT (Jian Xu, Lamouille, and Derynck 2009; Lamouille, Xu, and Derynck 2014; De Craene and Berx 2013)**.** The EMT process involves a network made up of signal transduction pathways, a group of transcription factors, and microRNAs that modulate these transcription factors. There is a significant cross-talk and autocrine/paracrine feedback between signaling pathways involved in EMT; for example, TGFβ forms an autocrine loop with SNAIL1 (Hector Peinado, Quintanilla, and Cano 2003; Gregory et al. 2011) and also activates the SHH and Wnt pathways (Zhang, Tian, and Xing 2016; Steinway et al. 2014). Figure 4A indicates an aggregated representation of the pathways and TFs involved in TGFβ-driven EMT, with a particular focus on the model of (Steinway et al. 2015, 2014), which focused on hepatocellular carcinoma. Note that nodes and edges in this meta-network might be functional in a context-specific (and even cell-specific) manner. A striking feature of this meta-network (also shared by Figure 1) is the existence of multiple positive feedbacks between signal transduction pathways and transcription factors. The transcription factor HEY1, a downstream effector of Notch signaling and part of the same family of TFs as TWIST, stands out due to its known role as an EMT inducer (Zavadil et al. 2004) but a lack of known feedbacks or interactions with other transcription factors (neither directly nor indirectly). Apart from HEY1, there are several TFs not included in Figure 4 but known to be effectors of signaling pathways that can induce EMT, such as FOXO3 and FOXA1 (Belguise, Guo, and Sonenshein 2007; Chou et al. 2014; Song, Washington, and Crawford 2010; Guttilla et al. 2012) and there are hints that some of their effect might be independent of the canonical EMT TF core. For example, FOXO3 can directly activate E-cadherin transcription (Chou et al. 2014) and FOXA1 can maintain E-cadherin expression even under TGFβ or SNAIL1 induction (Chou et al. 2014; Song, Washington, and Crawford 2010). Thus, these transcription factors are examples of TFs that, if left out of the TF core, would violate the assumption that the EMT TF core integrates the signals for EMT decision making.

What is the consequence of not including unknown TFs that are key to EMT decision making? According to the EMT model of (Steinway et al. 2015, 2014) not including these TFs results in missing important cellular contexts in which hybrid E/M can arise. Importantly, a careful look at the signaling pathways might allow us to identify these contexts even if the key TFs are unknown or left out. In (Steinway et al. 2015, 2014), SMAD KO gives rise to hybrid E/M states in which all but one of the EMT TFs are active, the exception being the TF HEY1 (Figure 4E). In contrast, most of the signaling pathways are in what would be expected from an E state. In particular, RAS, PI3K/AKT, NOTCH, SHH, and WNT pathways have low activity. Thus, the hybrid aspect of these states, defined by having most EMT TFs in their active state while simultaneously having HEY1 and multiple EMT-associated signaling pathways in their inactive state, is only apparent when looking at both the signaling pathways and the EMT TFs. Note that if we only looked at the EMT TFs or the EMT marker E-cadherin, the only clue about the "hybridness" of this state would have come from HEY1. This illustrates an important point and a word of caution against relying too strongly on the known EMT TF core when defining a hybrid E/M state. It is possible that other TFs that are downstream of the affected pathways could influence EMT independently of the known TF core (Figure 4F). This cautionary note also applies to wet-lab work, since the experimentally used E/M markers do not include noncanonical EMT TFs and reporters of EMT-associated signaling pathways (Jolly et al. 2018). Although the ideal solution would be to have a system-wide understanding of the EMT TFs and how they regulate the hallmark markers of EMT, the first main step towards addressing this problem should be to connect mathematical models with the data-based studies that have made progress in this direction (George et al. 2017; Liberzon et al. 2015; Tan et al. 2014).

**Network-level control of EMT and cancer metastasis**

Controlling complex systems is important to all aspects of human life. This includes biomedicine, where controlling complex molecular networks could lead to reprogramming diseased cells towards more normal physiological states. In the context of EMT, control could mean driving cells to the mesenchymal phenotype in the absence of EMT signals (Figure 4C). Conversely, control could mean driving cells that would otherwise undergo EMT to an epithelial phenotype.

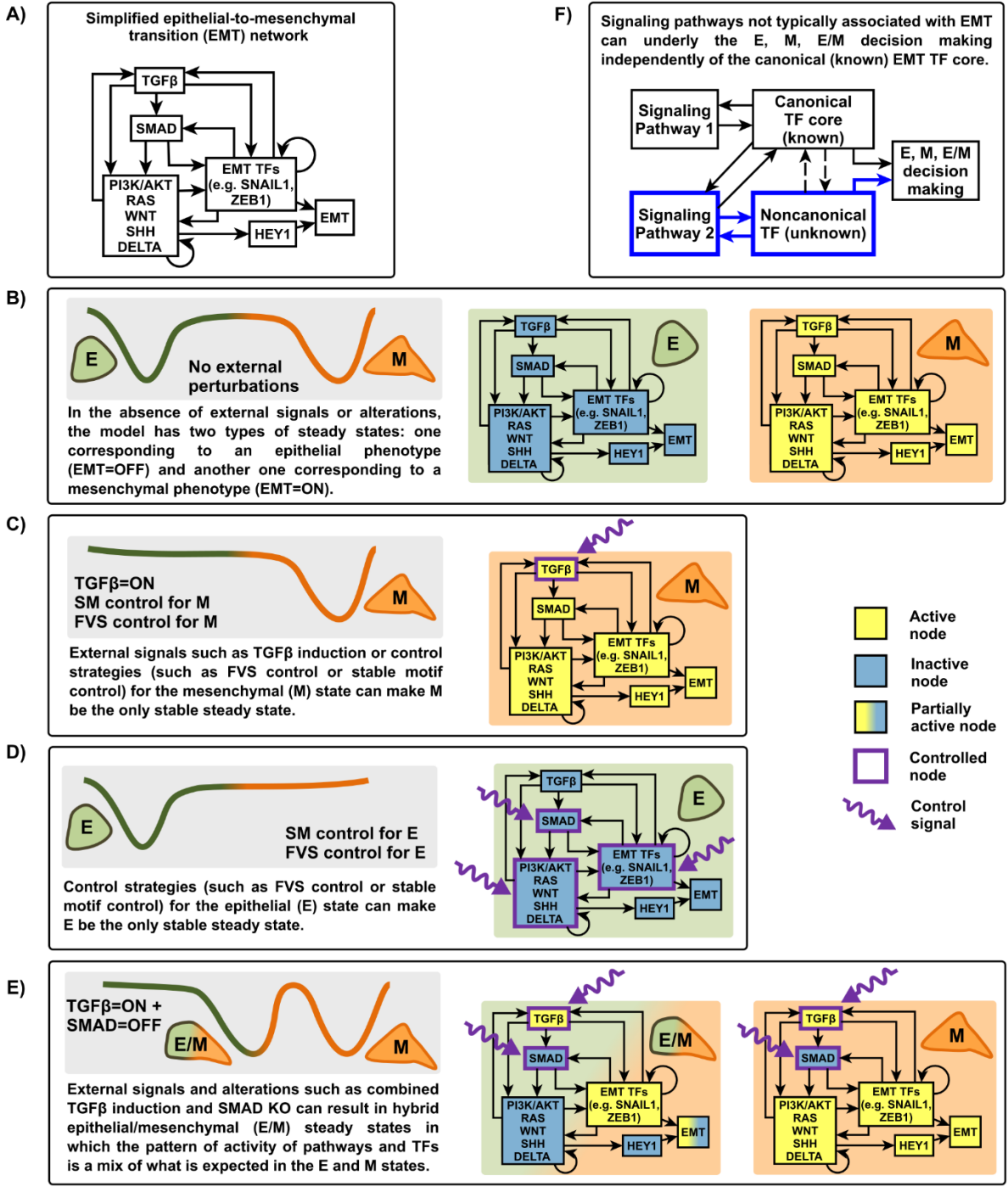

**Figure 4. Network modeling of the epithelial-to-mesenchymal transition (EMT).**
(A) Simplified version of the EMT network model of Steinway et al. The network includes multiple pathways, TFs, and crosstalks that are known to be involved in EMT and are used in multiple mathematical models reviewed here. (B) The model has an epithelial E and a mesenchymal M stable steady state in the absence of external signals. The colored valleys illustrate the attractor landscape. (C, D) External signals or attractor control interventions such as stable motif (SM) control or feedback vertex set (FVS) control, described in detail in the next section and Figure 5, can modify the attractor landscape so that only the E or M state is a stable steady state. (E) Hybrid epithelial/mesenchymal states arise in the model under certain perturbations (e.g. combined TGFβ induction and SMAD KO). (F) A signaling pathway can underlie the EMT decision making process through a noncanonical TF (blue edges and node outlines) and bypassing the canonical (known) TF core.

From a clinical cancer research perspective, targeting ways to inhibit cancer metastasis is an important focus because metastasis is the major cause of morbidity and mortality in patients with solid tumors (Steeg 2016). EMT is a critical process to cancer metastasis; therefore, identifying strategies for suppressing EMT remains crucial clinically. There exist drugs on the market and within clinical trials that target tyrosine kinases, which participate in the upstream signal transduction pathways involved in the EMT process (see Figure 1 and 4A). These upstream pathways are not specific to EMT; thus, targeting these pathways produces side effects on other cellular pathways controlling, for example, cellular proliferation and immune system evasion. For example, inhibitors of TGFβ have been shown to inhibit metastasis in pre-clinical models and are approved for clinical use (Mohammad et al. 2011; L. Yang 2010). However, clinically, TGFβ inhibitors have significant side effect profiles, perhaps related to their role in myriad cellular contexts (Costabel et al. 2014). Even at the cellular level, TGFβ has a dual role in tumor suppression and metastatic progression (L. Yang 2010; Jia Xu et al. 2015). The ideal clinical targets for EMT would be the transcription factors that target hallmarks of EMT such as E-cadherin expression; unfortunately, no such drugs exist at this time. In the absence of effective single-target drugs, it is important to determine what combinatorial therapies could be successful. There are two components of a therapeutic/control intervention: determining the targets and the control actions on these targets, and implementing the control actions. In the following sections we will consider the theoretical and practical aspects in turn.

*To what extent does the structure of the EMT network determine our ability to control it?*

Complex systems involve many parameters, most of which may be unknown. Thus, seeking parameter-independent approaches to network control is highly desirable. Some recent advances in network theory include structural control approaches that design control strategies for the dynamics of a network based solely on their topology. Feedback vertex set (FVS) control is a network control methodology that uses the structure of a network (its wiring diagram) and knowledge of biomarkers in the target cell fate of interest to reprogram a cell. In FVS control we manipulate the internal state of the FVS, a group of nodes that intersect every cycle in the network, which results in the disruption of all the feedback mechanisms and the ability to control the network. The defining characteristics of FVS control are (i) that it allows control of the network towards its naturally occurring states, (ii) that the control actions required are simple and consist of fixing the internal state of the FVS to match that of the target state of interest, and (iii) that it makes robust predictions that are not dependent on the details of the underlying dynamics of the network but only on the network and its interactions. When FVS control is applied to a regulatory network, the naturally occurring states correspond to cell fates and the required control actions correspond to manipulating the expression or activity level of TFs or signaling proteins to match those in the cell fate of interest. The effectiveness of FVS control interventions was demonstrated experimentally (Kobayashi et al. 2018).

In the case of the EMT network, FVS control would indicate the group of nodes that simultaneously satisfy two conditions: (i) their control to achieve their expression/activity level characteristic to epithelial cells would drive the whole network to the epithelial state and (ii) their control to achieve their expression/activity level in mesenchymal cells would drive the whole network to a mesenchymal state. If hybrid E/M states are considered, the group of FVS nodes would also satisfy a similar condition for each hybrid state.

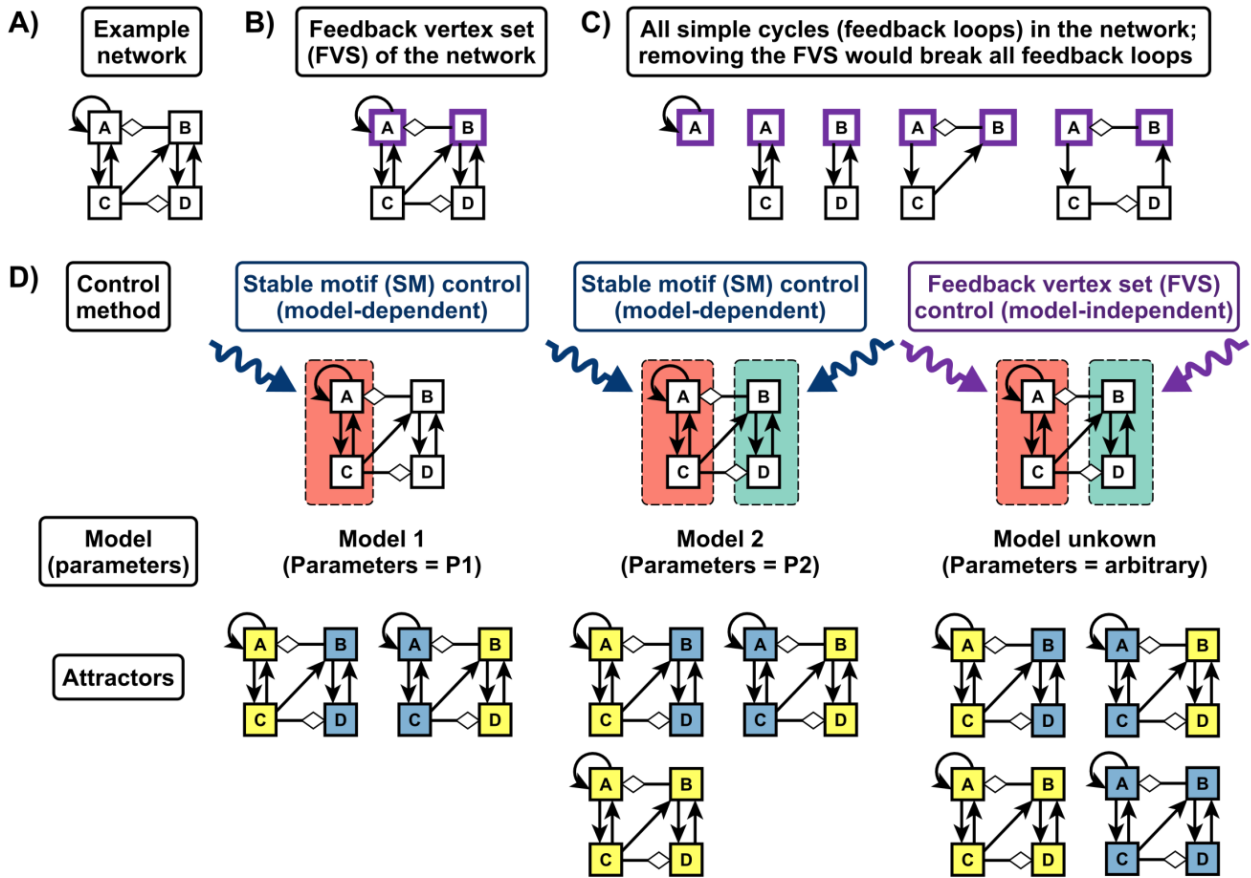

**Figure 5. Model-dependent and mode-independent control strategies using stable motif control and feedback vertex set control.**
Stable motif (SM) control and feedback vertex set (FVS) control are attractor-control strategies that can steer the dynamics of the system towards one of its dynamical attractors (e.g. steady states). (A) Example network topology underlying a continuous or discrete mathematical model. Due to the multiple feedback loops in this network, it supports multistability. (B-C) The FVS of a network (panel B) is a set of nodes that, if removed, break all feedback loops of the network (panel C). (D) SM control makes model-dependent predictions; it uses the parameters of a model to determine which feedback loops are sufficient for steering the model towards any of its attractors. FVS control makes model-independent predictions; it uses solely the topology of a model to determine nodes (the FVS) that can steer any model that shares the topology towards any if its attractors. In the representation of the attractors yellow stand for a high value of the state variable and blue indicates a low value of the state variable.

FVS control was shown to apply to both continuous and discrete systems. It is model-independent in the sense that if two models have the same wiring diagram, they have the same FVS. This is because the essence of FVS control is that controlling all cycles is sufficient to control the attractor of the system. This way the FVS represents the worst-case scenario (upper limit) of control targets. There is emerging evidence that in situations where a dynamic model can be specified and parameterized, one only needs to control certain cycles for each attractor, and thus targeting a subset of the FVS is enough (Zanudo and Albert 2015; Gan and Albert 2018; Jorge Gomez Tejeda Zañudo, Yang, and Albert 2017; Mochizuki et al. 2013; Rozum and Albert 2018).

Figure 5 illustrates this point in the case of an example network underlying a continuous or discrete mathematical model. There are five feedback loops in the network (Figure 5C), and its FVS contains two nodes (Figure 5B). Notably, two positive feedback loops (those formed by A and C, and by B and D, respectively), may be self-sustaining, depending on the specifics of their coupling to each other. Under different dynamic models of this network (or different parameterizations) its attractor repertoire will differ, as will the number of control targets whose control can drive the system into one of its attractors. In the scenario with parameter set P1 the BD loop always attains the opposite state of the AC loop, thus the system has two attractors, and control of A is sufficient to determine the attractor of the system. In the scenario with parameter set P2 there are three attractors and both A and B need to be controlled to drive the system to one of these attractors. Nodes A and B form the FVS of the system, thus their control ensures convergence to any of its four attractors.

*How quantitatively precise do the control interventions need to be?*

Considering the ubiquitous stochasticity of molecule levels in various cell types, it is likely that the cell's regulatory networks and the corresponding cellular states are robust to moderate molecule level fluctuations, but also may be reprogrammable by noisy control signals if these signals are sufficiently strong. In other words, cellular states can tolerate noise but can also obey noisy control signals. What precision is ultimately required for human control signals depends on the stability of cellular states and their inherent stochasticity, which remains to be determined.

In terms of FVS control, attractor control towards a desired attractor (which is assumed to be an attractor of the original system) can only be guaranteed if the nodes of the FVS are controlled to be in the exact state as their states in the desired attractor. Because the thus-controlled network is effectively acyclic, multi-stability (of either steady states or cyclic attractors) is not possible (see Figure 4C, D). In other words, a FVS-controlled network has a unique global attractor (Fiedler et al. 2013). A result of the acyclic structure of the FVS-controlled network is that imprecisely controlling the FVS will result in a similar attractor to the desired attractor, with an error between attractors that can be quantified using explicit recursive formulas (Fiedler et al. 2013).

The case of implementation of model-dependent control interventions is similar to the FVS control case; if the implementation is exact, the controlled network is guaranteed to have a single attractor. However, since the controlled system retains feedback loops, imprecise control can result in multi-stability, and the potential to reach an undesired attractor. This undesired attractor may be similar to one of the original system's attractors or it may be a newly created attractor.

Incomplete control (i.e. only implementing the control of a subset of the FVS or of a model-dependent control set) can create new attractors if the intervention, through inducing the stabilization of network components or disrupting paths, leads to the elimination of previous dependencies of the original system. For example, in the first model on Figure 4 the BD feedback loop's state depends on the AC feedback loop's state. If imprecise control of node A weakens the inhibition of D by C, the BD feedback loop may become independent and a new attractor where all four nodes have high state variables may be created. The analogy is that a strong signal on the original system causes each module of the network to "point in the same direction". When the system is perturbed, some of the modules are allowed to point in a different direction, which results in multi-stability.

Indeed, the creation of new attractors was observed in Steinway et al 2015. In this model, stable motif control interventions to drive the system to an epithelial state necessitate control of five nodes. Using subsets of the stable motif control interventions (e.g. only SMAD=OFF) created multiple new hybrid states. There was a similar effect for the case of the SMAD=OFF intervention in the presence of TGFβ (see Figure 4E). In both of these cases, the effect of the intervention was still in the expected direction as it resulted in a decrease in the number of states that can reach the mesenchymal state.

Recent work in model-dependent control approaches give promising indications that control interventions may not always need to be quantitatively precise. Stable motifs in continuous models involve inequalities in the state variables (or equivalently, regions of the state space). As long as the intervention causes (directly or indirectly) that the variables stay within the range of the inequality, then the intervention will succeed in trapping the system within a subspace (Rozum and Albert 2018). New attractors could still be created by the intervention but are restricted to the inequalities of the trap subspace. Further work could define more precise inequalities in the stable motifs that result in trap subspaces that contain only the attractor of interest.

Going back to the epithelial-to-mesenchymal transition, these control-theoretical inquiries raise the question of what exactly defines an E state, M state, or E/M state even in terms of well-known markers. Are they specific values of these markers or rather intervals? The good news is that synthetic biology technology now allows the development of detectors that sense the level of EMT regulators and counteract undesired states. We will describe these technologies next.

**Synthetic biological tools as detectors, modifiers and controllers of EMT**

*EMT detectors*

Before controlling gene expression levels in endogenous biological settings, we must know the actual levels of EMT/metastasis regulators, and how they deviate from normal ranges. EMT detectors sense the levels of EMT/metastasis-regulators. This requires the construction of genetic sensors or detectors that can process the information of intracellular levels in real time. Over the years, genetic tools such as biological decoders, logic gates, memory storage devices and feedback systems that allow real-time monitoring of markers have been constructed. These types of systems can lay the foundation for detecting endogenous disease signatures (Auslander et al., 2018, Guinn and Bleris, 2014, Weinberg et al., 2017, Farzadfard and Lu, 2014, Nevozhay et al., 2013, Kemmer et al., 2010), allowing greater specificity for specific biological markers, thereby enabling the detection of various cellular states. Such tools could be used to detect EMT induction to hybrid or fully mesenchymal states by matching phenotype and disease-signature states more precisely. For example, memory-based gene circuits could serve as lineage-tracing tools to observe reversibility in cellular phenotypes and multi-input logic gates could provide clues to phenotype switching in real time without the need for cell harvesting and destruction. This could be accomplished, for instance, by generating gene circuits that can respond to multiple markers such as transcription factors (e.g. ZEB1/SNAI1) and microRNA (e.g. miR200). Such a circuit could be hooked to a fluorescence reporter that would only generate an output (e.g. GFP) when there was high transcription factor concentration and low microRNA concentration (Figure 6A). In a population of cells, this could then be used to dynamically track individual cells and when cellular states are switched (Burrill et al. 2012; Al'Khafaji et al. 2018).

Since single targets may be insufficient for inducing mesenchymal states, eventual actuation of cellular responses requires multi-input detectors, for gaining better resolution for disease states. One solution with many manifestations over the past decade include 'digital sensors' which function as Boolean logic gates. Such tools can respond to multiple biological markers through various logical operations, yielding specific cellular responses only under desired conditions (Cho, Collins, and Wong 2018; Wong and Wong 2018; Roquet et al. 2016; Nissim et al. 2017; Siuti, Yazbek, and Lu 2013). Many EMT studies have focused on expressing markers of a single EMT/metastasis activator or inhibitor, which may be insufficient for detecting and controlling abnormal phenotypes. Utilizing tools that detect multiple markers may lay the groundwork for overcoming some of the limitations for systems detecting individual markers of EMT that have produced conflicting area of evidence (Figure 6B) (Fischer et al. 2015; Ye et al. 2017). Specifically, there have been several studies that have begun to lay the foundation for detecting multi-input EMT markers (Somarelli et al. 2013; Toneff et al. 2016), which has the potential to expand in the future with additional synthetic biology tools. Previous tools that can be used to move this work forward are proof-of-principle logic gates that can respond to various combinations of two inputs (e.g. AND, OR, NOR XOR, see Figure 6B). Such devices typically respond to inputs in a binary fashion and yield a single output, however such devices can be scaled to respond to multiple biological markers.

Genetic tools with binary outputs are useful in synthetic biology, but they may be insufficient for producing signals more in alignment with endogenous biology, such as analog signals. Building genetic tools responding in analog fashion would be extremely valuable to the area of EMT. For example, it would be useful to design an analog gene circuit that can control pre-defined levels of EMT/metastasis activators or inhibitors could help observe more precisely the conditions where EMT or other metastatic steps occur, and to what extent hybrid cells exist in a population (Figure 6C). Gene circuits potentially adaptable for this purpose have been previously constructed, allowing tunable doses of specific gene products (Nevozhay, Zal, and Balazsi 2013; Wang, Barahona, and Buck 2014). Cancer biology could benefit from such tools that enable controlling and elucidating new information on EMT both *in vitro* and *in vivo.* Such analog circuits could function by integrating a feedback mechanism, allowing the creation of dose-responsive characteristics. The circuit could then respond to a particular stimulus (e.g. chemical) to weaken repression and feedback in the system such that a gene-of-interest (GOI) can increasingly be expressed. In the case of EMT, a GOI can be a protein-coding gene like ZEB1, where dose-response characteristics can be studied to observe threshold changes that occur under different levels of induction. More generally, these feedback systems could provide the platform for precisely detecting and possibly regulating the levels of EMT/metastasis genes of interest in a wide dynamic range to study cellular transition thresholds (Figure 6C) by converting gene circuit stimulus precisely into desired gene expression levels. These circuit detectors would provide a working model for eventual expansion into detectors of endogenous stimuli in place of synthetic stimuli controlling gene expression levels or allow actuation in the form of feedback into endogenous EMT biology.

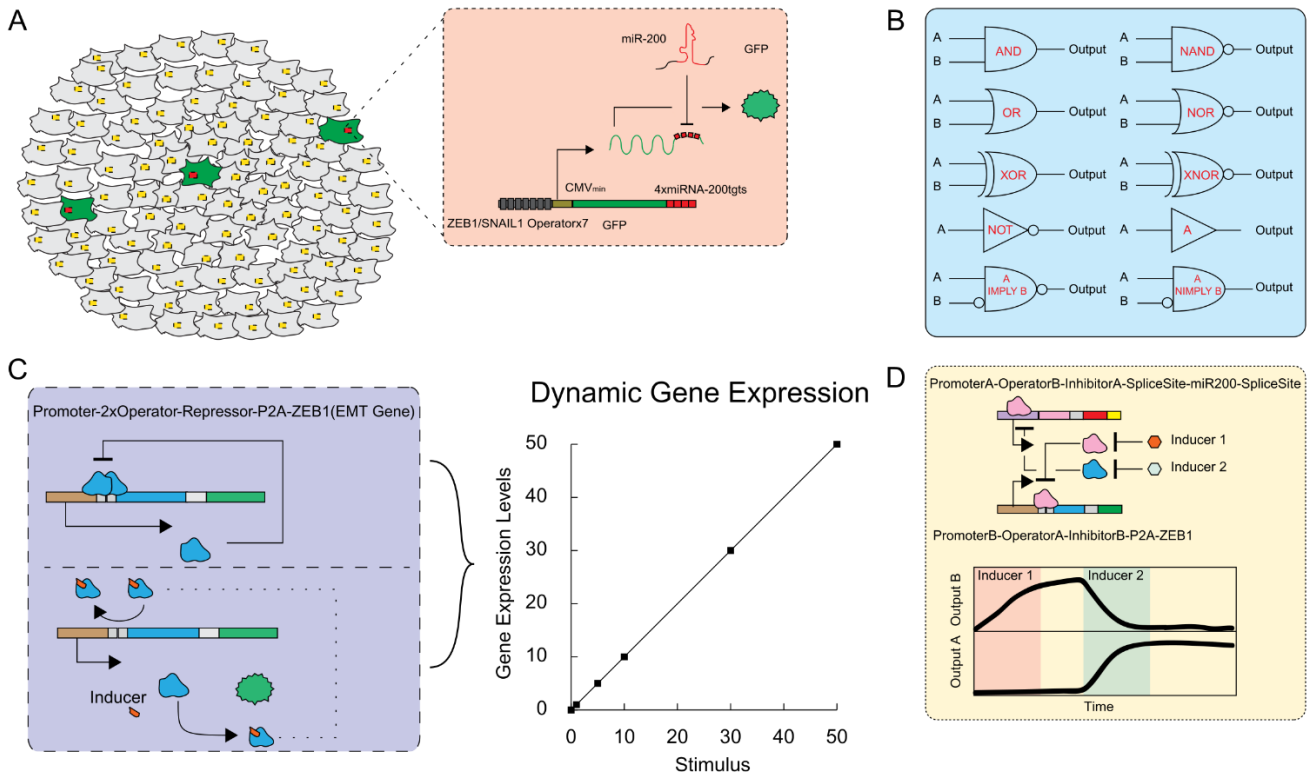

**Figure 6. Synthetic Biology Detectors.**
(A) Schematic illustration of gene circuit integrated in a cell population of interest allowing lineage tracking of cellular profiles among a population of heterogeneous cell types that are isogenic. (B) Schematic illustrations for higher order gene circuits that could be recruited to serve as sensors for multiple EMT markers to initiate detection and actuation in real time. (C) Schematic illustration of an example gene circuit that could allow robust and wide dynamic control of gene expression over a wide range of stimulus to juxtapose ON/OFF states of overexpression/knockout. (D) Schematic illustration of a synthetic biology gene circuit that can be used to observe controlled switching between epithelial and mesenchymal states by overexpressing targets that promote EMT in one state and targets that inhibit EMT in the alternative state.

A key aspect that has been elucidated regarding the EMT process is the existence of two cross-talking toggle switches between SNAIL1 & miR-34 family on one end and ZEB1/2 and miR-200 family on the other. As one of the premier papers that launched the field of synthetic biology, Gardner et al. created a genetic toggle switch in bacteria (Gardner, Cantor, and Collins 2000). This launched the engineering disciplines that have worked for over a decade to create novel circuits and tools that can interface with endogenous biology. Since the original paper, toggle switches have been widely characterized with methods for control, tunability, and precision (Y. Xu et al. 2016; Lebar et al. 2014; Kramer et al. 2004). Such work has also allowed insight into what sort of dynamics and kinetic parameters can be manipulated within the core EMT network. Designing controlled systems to function as toggle switch motifs and mimic natural EMT circuitry may allow elucidation of thresholds which, when passed, serve as crucial transition points between epithelial and mesenchymal states (Figure 6D). These switching systems would function by producing two mutually inhibitory repressors which act on each other through modifying operator sites between promoters and the GOI. Specific EMT genes can then be produced from each node of the toggle switch (e.g. ZEB1 or miR200) that can flip ON or OFF depending on which state the toggle switch is in. This can easily be controlled by chemical stimuli that regulate the function of the repressors. When one chemical is added, the toggle switch can flip and begin expressing one EMT gene. By adding the second chemical, the other EMT gene can be expressed, allowing the study of rapidly switching EMT states. These toggle switch detectors, like the feedback systems, could interface with endogenous inducers, initially serving as sensors, but with further modifications be expanded into actuators, which we will discuss now.

*EMT actuators*

EMT actuators provide fine, reversible, time-dependent, user-defined, and cell-autonomous control over the EMT network. They can be considered expansions of biological detectors and modifiers for controlling EMT biology. Synthetic gene circuits purposed as actuators build upon the foundation of genetic sensors or detectors of endogenous information, but in place of fluorescence or luminescent outputs, they provide a functional response to counteract certain undesired cellular states. For example, many of the biological gates described previously serve as proof-of-principle devices with outputs that are measurable, e.g., through fluorescence intensity. Such systems can be repurposed to produce functional genes ranging from apoptotic triggers to cell reprogramming genes.

For example, graded feedback circuits that sense endogenous EMT transcription factors could induce progressively higher levels of counter-acting proteins when EMT inducers become too high, and thus maintain the epithelial state, as has been done with various cancer sensors (Ehrhardt et al. 2015; Xie et al. 2011). Such sensors have been constructed for transcription factors for processes involved in other disease models and given the ease to which such systems can be constructed and multiplexed to find optimal gene expression, various sensors could be created for the main EMT regulators (Figure 7A). In such a scenario, the genetic detectors outlined in the previous section could be hooked up to functional genes in a library-like fashion, allowing many different gene circuit variants that respond to EMT markers. The same genetic architectures as in the sensors could be utilized, to control any desired functional gene (e.g. apoptotic genes, proliferation genes, and stem-cell reprogramming genes). Simple sensors that are re-engineered into single-response actuators could further be expanded into higher order actuators by utilizing tools such as biological decoders, logic gates, memory storage devices and feedback systems (Figure 7B) (Auslander et al. 2018; Guinn and Bleris 2014; B. H. Weinberg et al. 2017; Farzadfard and Lu 2014; Nevozhay, Zal, and Balazsi 2013; Kemmer et al. 2010). Such a gene circuit actuator would allow refining the specificity for EMT states by incorporating several markers as stimuli, but also allow precise classification under different states. For example, in the case of the decoder (Figure 7B), a state with three inputs can have up to eight different combinations. In theory, this would allow an actuator with eight different responses depending on the levels of each individual marker. Some states may be more heavily mesenchymal while others more epithelial with others remaining intermediate. Creating actuators with multiple outputs could increase the versatility and strength of tools for probing EMT biology, allowing varying levels of interventions depending on the input combinations the circuit receives. Expanding the range of single-input actuators into multi-input/output gene circuits could allow studying multi-node phenotypes that depend on multiple cellular markers (e.g. detecting high ZEB1, high SNAIL1, and intermediate miR-200 would lead to producing a high dose of EMT inhibitors).

Another avenue of potential EMT intervention coupled with circuit actuation is optogenetic tools that allow the capability to control single cells in a population of many cells (Figure 7C). Optogenetic tools could replace existing gene circuits (e.g. Linearizers) for expressing GOIs (e.g. ZEB1). This could allow single-cell level gene expression of EMT genes in cell-culture, currently impossible with chemical means. Such precise levels of control and monitoring could allow direct *in vitro* observation of individual cells undergoing EMT among a population of cells that are not undergoing EMT. Although less transferable to

*in vivo* systems, optogenetics has the potential to determine *in vitro* what levels of EMT inducers and inhibitors are necessary to convert hybrid or mesenchymal cell states to the epithelial phenotype or vice versa. Another area where synthetic biology can offer actuation or control of cells that have undergone EMT is the area of immunoengineering (Nissim et al. 2017). Recently, Nissim et al. created an immunomodulatory gene circuit platform for activating T cell-mediated killing of cancer cells. Such induction could be manipulated to create gene circuits that respond to certain EMT markers in circulating tumor cells (Figure 7D). This gene circuit actuator would allow therapeutic interventions once EMT has already occurred. In such a scenario, engineered mesenchymal cells could have gene circuits that produce membrane proteins that stimulate the immune system when the appropriate EMT genes are active. Consequently, engineered immune T cells would be recruited to produce 'destruction' proteins that lead to the eradication of rogue mesenchymal cells, therefore suppressing this cell type under improper conditions.

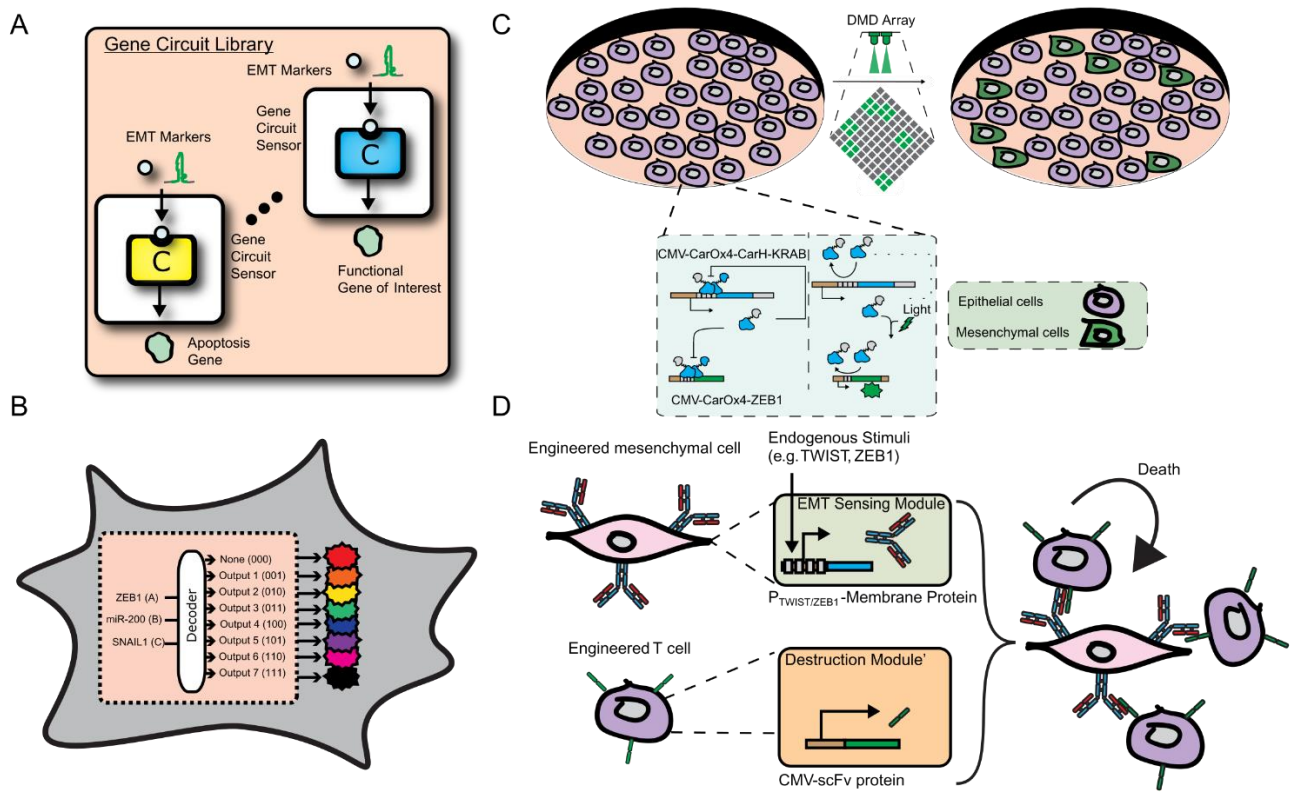

**Figure 7. Synthetic Biology Actuators.**
(A) Schematic illustration of genetic circuit library that could serve as sensors for various EMT genes in singlet and respond with appropriate response output (e.g. inducing apoptosis genes). (B) Schematic illustration of a synthetic biology gene circuit such as a decoder that could analyze the spectrum of cellular states relating to EMT to contrast classifications of single markers. (C) Schematic illustration of gene circuit controlling spatial induction of EMT to study how spatial parameters affect the development of transition between epithelial and mesenchymal cell types. Such a system could be accomplished with a digital mirror device (DMD). (D) Schematic illustration of immunoengineering as an outlet for post- EMT control by killing of mesenchymal cells by engineered T cells.

**Concluding remarks**

The epithelial-to-mesenchymal transition (EMT) in development, wound healing, and cancer seems to have a shared regulatory network of transcription factors. These transcription factors (TFs) interact with each other, with microRNAs, and with signal transductions pathways, forming numerous feedback loops. Despite the many advances at both the experimental and theoretical levels, there is still more to be learned about how the physical microenvironment modulates EMT, the intermediary EMT states, and the role of EMT and cancer metastasis. In particular, we discussed how intermediate EMT states can arise as a consequence of multiple network motifs (ternary switches, coupled bistable switches, and stable motifs) and through distinct biological mechanisms (microRNA-TF feedbacks or TF-signaling pathway feedbacks).

There is increasing evidence that interventions to block EMT and metastasis in cancer would need to be combinatorial because of the complexity of the underlying signaling and regulatory network. Theories of model-independent and model-dependent control like feedback vertex set (FVS) and stable motif control can guide the selection of targets. Recent exciting developments in synthetic biology allow the implementation of such combinatorial control. More work is needed to bring the theory and practice in synchrony; for example, it is still an open question how to best identify control sets with a smaller size than the FVS when only the structure of the network and limited information on the expected states and regulatory logic is available. But overall, the path is now open towards control of the regulatory networks that underlie EMT and metastasis. Theory-guided synthetic biology control methods could advance personalized cancer treatment by adjusting gene expression according to each tumor's metastatic tendencies.

**Funding**

Discrete dynamic modeling of cancer-related networks by J.G.T.Z. and R.A. is supported by the National Science Foundation (grant PHY 1545832), the Stand Up to Cancer Foundation, and a Stand Up to Cancer Foundation/The V Foundation Convergence Scholar Award (D2015-039) to J.G.T.Z.. R.A. is additionally funded by National Science Foundation grants MCB1715826 and IIS1814405. G.B. is funded by a NIH/NIGMS MIRA grant R35GM122561, and by the Laufer Center for Physical and Quantitative Biology. M.T.G. is supported by the NIH grant T32 GM008444.

**Acknowledgements**

J.G.T.Z. would like to thank Nikhil Wagle for hosting him as a Research Associate at the Dana-Farber Cancer Institute and an Associated Researcher at the Broad Institute of Harvard and MIT.

Phenotype: The Contribution of Multiple Integrated Integrin Receptors." *Journal of Mammary Gland Biology and Neoplasia* 3 (2): 191–200.